\documentclass[twocolumn,twocolappendix]{aastex701}

\usepackage{float}
\usepackage{placeins}
\usepackage{xcolor}

\begin{document}

\title{Spatial Property of Multiple Metallic Populations in the Tidal Stream of $\omega$ Centauri}

\newcommand{\SYSU}{Department of Astronomy, School of Physics and Astronomy, Sun Yat-sen University, Zhuhai, Guangdong Province, China}
\newcommand{\CSST}{CSST Science Center for the Guangdong-Hong Kong-Macau Greater Bay Area, Zhuhai 519082, China}
\newcommand{\UCN}{Instituto de Astronom\'ia, Universidad Cat\'olica del Norte, Av. Angamos 0610, Antofagasta, Chile}

\correspondingauthor{Baitian Tang}
\email{tangbt@mail.sysu.edu.cn}

\author{Shiru Zheng}
\email{zhengshr7@mail2.sysu.edu.cn} 
\affiliation{\SYSU}
\affiliation{\CSST}

\author{Baitian Tang}
\email{tangbt@mail.sysu.edu.cn}
\affiliation{\SYSU}
\affiliation{\CSST}

\author{Long Wang}
\email{wanglong8@mail.sysu.edu.cn} 
\affiliation{\SYSU}
\affiliation{\CSST}

\author{Jose G. Fernandez-Trincado}
\email{jose.fernandez@ucn.cl}
\affiliation{\UCN}

\author{Ruoyun Huang}
\email{huangry8@mail2.sysu.edu.cn}
\affiliation{\SYSU}
\affiliation{\CSST}

\author{Xia Li}
\email{lixia76@mail2.sysu.edu.cn}
\affiliation{\SYSU}
\affiliation{\CSST}

\author{XiaoDong Li}
\email{lixiaod25@mail.sysu.edu.cn}
\affiliation{\SYSU}
\affiliation{\CSST}

\begin{abstract}

$\omega$ Centauri, the remnant nucleus of an accreted dwarf galaxy, is a unique laboratory for studying complex stellar populations. The recently discovered Fimbulthul stream provides a fossil record of its ongoing tidal dissolution. In this work, we investigate the spatial distributions of metal-rich and metal-poor populations within $\omega$ Centauri and its stream to constrain the cluster's formation history. Using synthetic photometry from Gaia DR3 XP spectra, we classify stars via a Support Vector Classifier (SVC). The spatial distributions are then compared to a scaling N-body simulation performed with the PeTar code. Our analysis reveals no significant radial gradient in population ratios within the cluster, though the metal-rich stars may be slightly more extended. The population ratio in the tidal stream is consistent with that of the present-day cluster, albeit with large uncertainties. Our simulation indicates that any initial radial gradient must have been shallow, with a maximum fraction difference less than 0.15. Both observational and dynamical results suggest that the metal-rich population is not formed centrally concentrated. By combining our results and existing literature, we propose a new formation scenario for $\omega$ Centauri.

\end{abstract}



\section{Introduction}

According to the $\Lambda$ Cold Dark Matter ($\Lambda$CDM) model, the formation and evolution of the Milky Way (MW) have been significantly shaped by the accretion of dwarf galaxies. Several such merger events have been identified, including the Gaia–Sausage–Enceladus (GSE), Sequoia, Thamnos, and Helmi streams \citep[e.g.,][]{koppelman2019multiple, helmi2018merger, 2018MNRAS.478..611B, 2019MNRAS.486.3180K}.
The GSE event, one of the most significant, is estimated to have occurred $\sim10$ Gyr ago \citep[e.g.,][]{helmi2018merger, 2010A&A...511L..10N, 2018MNRAS.478..611B, 2018A&A...618A..78H} and largely disrupted the dwarf galaxy. The tidal debris from this merger has been mapped using precise Gaia astrometry and spectroscopic radial velocities (RVs), which reveal distinct structures in integrals-of-motion space \citep{Koppelman_2018, koppelman2019multiple}.
In this hierarchical framework, globular clusters (GCs) in the MW likely formed through two channels:
(1) in-situ formation, where clusters formed early within the proto-Galaxy;
(2) accretion from dwarf galaxies, where clusters formed externally and were subsequently captured. Notably, the accreted population includes nuclear star clusters (NSCs)—the remnant cores of dissolved dwarf galaxies.
Although these populations are now dynamically mixed, their distinct origins can often be traced through orbital dynamics \citep[e.g.,][]{myeong2018sausage, massari2019origin}, but MW accretion history may complicate such dynamical interpretations \citep[e.g.,][]{2025arXiv251013990P, 2023A&A...673A..86P, 2024A&A...690A.136M}. In comparison, detailed chemical patterns may provide the most definitive diagnostic of their origins \citep[e.g.,][]{LinTang2025}. 
Traditionally, star clusters were considered simple stellar populations, forming in a single event where all members share a common age and chemical composition. However, observations over the past few decades have revealed that most GCs host multiple stellar populations (MPs) with distinct chemical abundances \citep{2018ARA&A..56...83B, 2019A&ARv..27....8G, milone2022multiple, HuangTang2024, Tang2017,Tang2018}. These chemical variations manifest as split sequences in the UV-related color–magnitude diagram (CMD), observable across different evolutionary phases \citep[e.g.,][]{piotto2012hubble,milone2019hst,marino2008spectroscopic}. 
A powerful tool for identifying MPs is the ``chromosome map'' \citep{milone2017hubble},
a pseudo two-color diagram constructed from HST filters (F275W, F336W, F438W and F814W). In most GCs (Type I clusters), stars clearly separate into two groups: a first generation (FG) and a second generation (SG). However, in some GCs (Type II clusters) the FG and/or SG sequences appear to be split, exhibiting more complex patterns. Interestingly, most Type II clusters display variations in overall $C+N+O$ content, or heavy element \citep[e.g.,][]{milone2022multiple}. 
$\omega$ Centauri (NGC 5139) is a prototypical Type II cluster, characterized by a significant spread in iron abundance. More intriguingly, distinct light-element anti-correlations are observed even within its individual metallicity groups \citep{gratton2011multiple,milone2020chromosome, 2021MNRAS.505.1645M}.

The spatial and kinematic distributions of $\omega$ Centauri’s stellar populations provide critical constraints on its origin and dynamical evolution. A central concentration of SG (metal-rich) stars relative to FG (metal-poor) ones would support scenarios where SG (metal-rich) stars formed deeper in the potential well, while dynamical evolution (e.g., mass loss, relaxation) may gradually erase such differences \citep{Vesperini2013}.
The observed chemical abundances and spatial distributions of stars in $\omega$ Centauri have led to several proposed formation scenarios. These include: (1) a merger origin, where the cluster resulted from the merging of distinct star clusters or the nucleus of a dwarf galaxy\citep{1995ApJ...447..680N}. This is supported by the more extended spatial distribution of metal-rich red giant branch (RGB) stars compared to their metal-poor counterparts \citep{2020ApJ...891..167C}. Some spectroscopic studies also indicate full spatial mixing  between metal-rich and metal-poor populations in the core region($\sim 5$ arcmin)\citep{Nitschai2024oMEGACatIM}; (2) prolonged in-situ formation involving the inspiral of at least one globular cluster \citep{mason2025chemical, jofre2025studying}; and (3) a self-enrichment model, where the cluster formed from chemically inhomogeneous gas, as evidenced by the central concentration of populations enriched in p-capture elements \citep{2025arXiv250916719D, clontz2025}. This is also supported by the more centrally concentrated helium-enhanced blue main-sequence (bMS) stars \citep{2009A&A...507.1393B, 2024A&A...688A.180S} and white dwarfs \citep{2025arXiv251008715S}.

Furthermore, $\omega$ Centauri’s extensive tidal mass loss has produced identifiable tidal streams, such as the Fimbulthul stream \citep{ibata2019identification, ibata2021charting} and other structures extending up to $4^{\circ}$ \citep{2025MNRAS.537.2752K}. However, the limited spectroscopy of stream stars hampers detailed population tagging beyond the cluster’s tidal radius($r_{\rm t}$).

In this work, we analyze the spatial distributions of metal-poor and metal-rich populations within both the cluster and its tidal stream, combining the analysis with N-body simulation to further disentangle dynamical evolution of multiple metallic population of $\omega$ Centauri and its tidal stream, which could help us reveal the early formation history of $\omega$ Centauri.
We address several primary questions in detail: \textit{(1) How does the fraction of metal-poor population vary across the cluster and in the stream?  (2) Does the extended stream retain imprints of the early dynamical evolution of the cluster's multiple metallic populations? (3) Can we recover the initial spatial distribution of multiple metallic populations?} This paper is structured as follows. We begin by using Gaia XP synthetic photometry to separate member stars (in both the cluster and the stream) into metal-poor and metal-rich populations (Section \ref{sect:method}). We then analyze the spatial distributions of these populations (Section \ref{sect:morp}). By comparing our results with simulated clusters and tidal streams, we investigate the initial spatial distributions and the mechanisms of stellar escape (Section \ref{sect:sim}). Finally, we provide a discussion in Section \ref{sect:dis} and summary in Section \ref{sect:sum}.

\section{Data and Methodology}
\label{sect:method}
\subsection{Gaia XP Spectra}
Gaia DR3 provides low-resolution spectra ($R \sim 30-100$, $330<\lambda<1050$ nm) for approximately 200 million astronomical sources with G magnitude brighter than 17.5. Each spectrum, known as the XP spectrum, comprises two distinct spectra obtained by the two components of the Gaia spectro-photometer (referred to as BP and RP) \citep{de2023gaia, gaia2023gaia}. 
The low-resolution spectra are released through the Gaia archive in the form of two sets of 55 Hermite function coefficients, one for each channel (BP/RP). The externally calibrated spectra (flux versus wavelength) can be reconstructed from these coefficients using the public Python library GaiaXPy\footnote{https://gaia-dpci.github.io/GaiaXPy-website/}. Furthermore, this tool allows for the calculation of synthetic magnitudes by convolving the reconstructed spectra with different filter transmission curves.


\begin{figure*}
\centering
\includegraphics[width=0.8\textwidth]{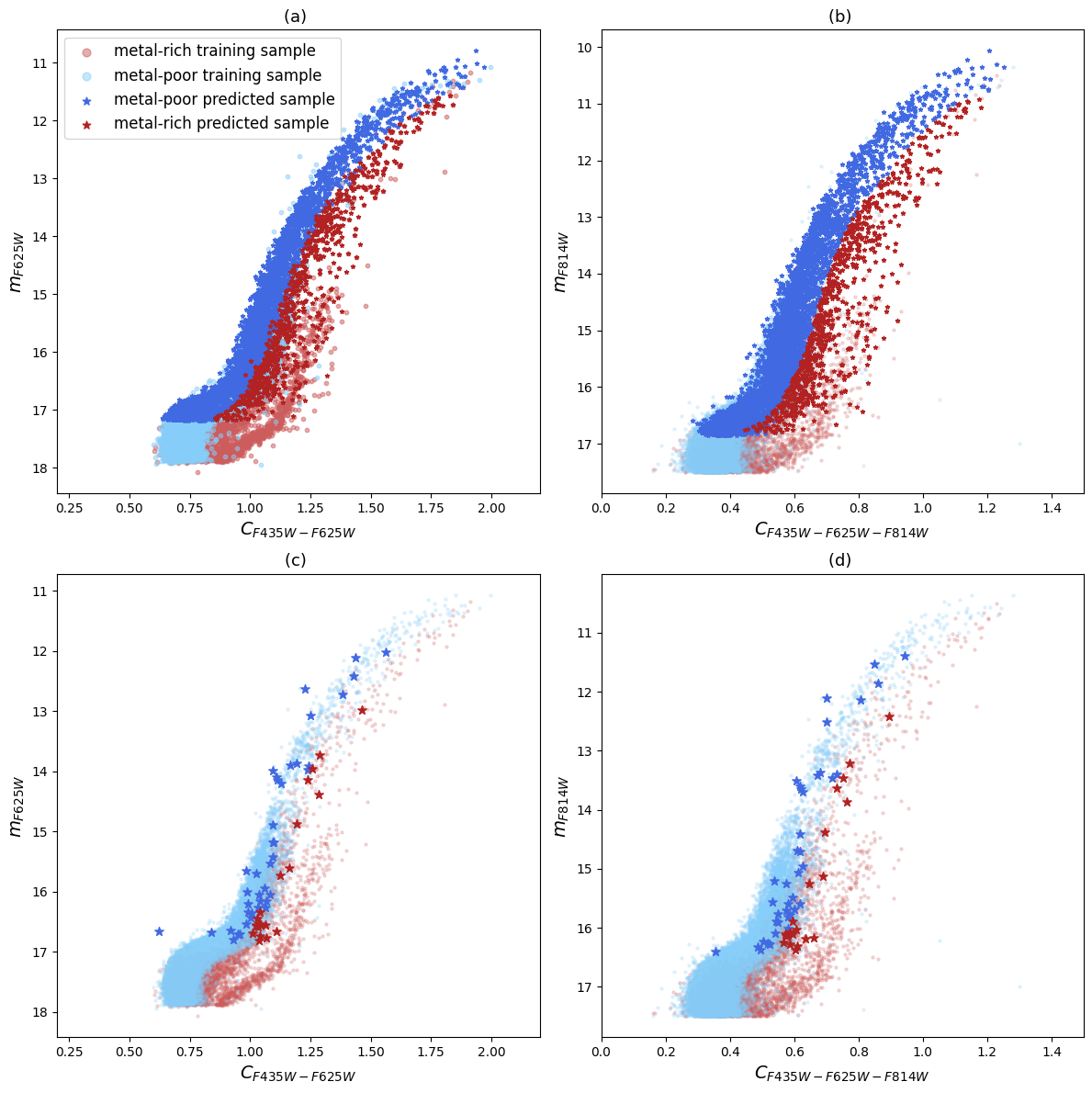}
\caption{Color–magnitude diagram and pseudo–color–magnitude diagram of the training and predicted stars. The light–red and light–blue dots represent the metal–rich and metal–poor stars in the training sample, respectively. In the two upper panels, the cluster members are separated into metal–poor and metal–rich populations, indicated by blue and red stars, respectively. In the two lower panels, the stream members are likewise separated into metal–poor and metal–rich populations, marked by blue and red stars. }
\label{fig:class}
\end{figure*}

\subsection{Member Selection}

\textbf{Stream members}: After cross-matching the Gaia XP database with the catalogs of stream member candidates in \citet{ibata2024charting}, we obtained the RA, Dec, proper motion, parallax, BP, RP, and G band magnitudes for 763 stars in the Fimbulthul stream (Figure \ref{fig:pos}a).
To characterize their metallicity, we further calculated their HST F435W, F625W, and F814W magnitudes using Gaia XPy(as was done for cluster members). We select stream stars brighter than $F625W = 17$, yielding a sample of about 40 stars to ensure reliable population separation. The distances were adopted from \citet{bailer2021estimating}.
Finally, we corrected the Galactic Dust Reddening and Extinction\footnote{https://irsa.ipac.caltech.edu/applications/DUST/} for each star.

\vspace{0.2cm}
\noindent\textbf{Cluster members} were selected from Gaia DR3 based on three criteria: 
(1) inside the cluster tidal radius (0.8$^\circ$);
(2) proper motion within cluster mean $\pm 1\sigma$;
(3) G $<$ 17.5. 
They are further refined by eliminating stars with large photometric uncertainties following the method described in APPENDIX. These cluster stars are shown in Figure \ref{fig:pos}(b).



\subsection{Separating metal-poor and metal-rich Populations}

To separate the stellar populations in both the cluster and the stream, we used two colors:  $C_{{F435W-F625W}}$, $C_{{F435W-F625W-F814W}}$\footnote{ ($m_{{F435W}}$- $m_{{F625W}}$)-($m_{{F625W}}$-$m_{{F814W}}$)}. \citet{Nitschai2024oMEGACatIM} showed that these two colors are able to effectively separate stellar populations with various metallicities.
To classify stellar populations into metal-poor or metal-rich, we applied the SVM (support vector machine) classifier algorithm\footnote{https://scikit-learn.org/stable/modules/svm.html\#svm-classification}. The training samples consist of two catalogs: (1) metallicities derived from MUSE spectra \citep{Nitschai2024oMEGACatIM}. These stars were observed by HST and located in the cluster center (Figure \ref{fig:pos}b). They cover the lower RGB\footnote{Stars in the upper RGB are saturated.}, SGB and upper MS. (2) accurate metallicities derived from high-resolution APOGEE spectra \citep{2021MNRAS.505.1645M}. These stars are mostly RGB stars brighter than 14 (Figure \ref{fig:pos}b). Their HST filter magnitudes are also derived from their Gaia XP spectra.

The procedures of classification can be summarized into three steps: (A) separating the stars in the training sample into two populations with the metallicity limit: [Fe/H$]=-1.4$. The metal-poor (metal-rich) population is labeled as class ``1'' (``2'');
(B) a hyperplane was constructed in a four-dimensional space defined by  $C_{{F435W-F625W}}$, $m_{{F625W}}$, $C_{{F435W-F625W-F814W}}$, $m_{{F814W}}$, which separates the two classes by maximizing the margin between their closest samples; (C) this hyperplane was used to predict the class membership of stars with unknown metallicities in the cluster and stream. 
The classification results are presented in Fig.~\ref{fig:class}.


\section{Spatial Distributions of Multiple Metallic Populations}
\label{sect:morp}

To investigate the radial distributions of the two populations, we estimated the fraction of stars belonging to each population within bins defined by radial distance from the cluster center. Each bin contains the same number of stars: 1000 stars per bin for the cluster and 13 stars per bin for the stream. The radial distance of a star in the stream is defined relative to the nearest position along the stream track, as fitted by \citet{mateu2023galstreams}.

\textbf{Cluster stars}: Our sample covers a wide radial range, extending from 5 arcmin (the half-light radius) to 48 arcmin ($r_{\rm t}$). The blackblue line of Figure~\ref{fig:sim_radial}  shows a flat trend in the fraction of metal-poor stars with increasing radius, when consider the uncertainty. The uncertainties of the fraction of metal-poor populations are calculated by adopting a Gaussian Monte Carlo (MC) approach. Within this region (5–48 arcmin), the global fraction of the metal-poor population is about 0.85, which is similar to the fraction of stars with metallicities below –1.4 reported by \citet{2021MNRAS.505.1645M}. 
In the core region, previous studies have reported no significant metallicity gradient within the half-light radius ($\sim 5$ arcmin), and the different sub-populations appear to be well mixed \citep{Nitschai2024oMEGACatIM}. While more extended radial distribution are reported at larger radii ($\sim 20$ arcmin)for metal-rich population\citep{2020ApJ...891..167C}. Furthermore, \citet{2025MNRAS.537.2752K} also suggested that the two populations are well mixed out to 4 degrees from the cluster center.

\textbf{Stream stars}: As shown in Figure~\ref{fig:sim_radial}, the fraction of metal-poor stars decreases with increasing radius. Overall, the fraction of the metal-poor population in the stream appears to be lower than that in the cluster. However, when the relative uncertainties are taken into account, the metal-poor fraction in the stream may still be consistent with that of the cluster.Unlike cluster stars, stream stars have different heliocentric distances, which inevitably introduces errors when calculating their apparent magnitudes at the cluster’s heliocentric distance. This  propagates to the fraction uncertainties of metal-poor populations.

We assess the potential bias introduced by CMD-based selection. The CMD-selected stream subsample contains only $\sim$40 stars brighter than 17 in the F625W band and therefore has a larger fractional uncertainty than the full stream sample ($\sim$760 stars). Although the metal-poor fraction inferred from the full sample is closer to that of the cluster, most stream stars occupy the faint region of the CMD, where different populations overlap and cannot be reliably separated in either the CMD or pseudo-CMD. Consequently, we preferentially adopt the results based on the CMD-selected subsample, for which a reliable population classification is possible.

Regarding completeness, our analysis is limited to sources brighter than $G = 17.5$. 
Based on the Gaia DR3 selection function \citep{CantatGaudin2023}, the catalog is essentially complete ($>99\%$) for $G \lesssim 18$ over most of the sky. 
Since neither the stream nor the analyzed cluster regions reach source densities where Gaia completeness is significantly affected, incompleteness is expected to have a negligible impact on our population ratios.

We also tested several other metallicity limits other than [Fe/H$]=-1.4$ (e.g., $-1.5, -1.3$), and explored alternative color indices (such as $C_{B-I}$ and $C_{B-V-I}$). In all cases, we obtain consistent radial distributions of different populations.

\section{Simulation of the dynamical evolution of Multiple metallic populations}
\label{sect:sim}
We simulated the dynamical evolution of $\omega$ Centauri  to reveal the initial spatial distributions of MPs and explore the influence of the dynamical evolution on its different populations. In this work, we used the state-of-the-art N-body simulation code, PeTar \citep{Wang2020petarAH}, which is particularly suitable for studying dense star clusters, binary interactions, and tidal disruption processes. The code can accurately handle an arbitrary fraction of multiple systems (e.g. binaries, triples) while keeping a high performance by using the hybrid parallelization methods with MPI, OpenMP, SIMD instructions and GPU, by combining the methods of Barnes-Hut tree, Hermite integrator and slow-down algorithmic regularization (SDAR). 
\subsection{scaling model of $\omega$ Centauri  }
$\omega$ Centauri is a massive star cluster with estimated mass of $ \sim 3.55 \times 10^6M_\odot$ and half mass radiu $r_h$ =10pc \citep{2018MNRAS.478.1520B}, it costs great computational resources to do a 1:1 modeling of its stellar evolution and dynamical evolution over a time interval of about 12 Gyr. 
As a discussion of the implication of our observations, we performed a scaled-down simulation with a total mass of $M = 1.04 \times 10^5M_\odot$, particle number of $N=3\times 10^5$, and the half mass radius of $r_h$ = 10 pc. 
Owing to the mass difference, the simulated cluster's dynamical timescales differ from the real cluster. However, the stream morphology and kinematics are less sensitive to internal dynamics than to the Galactic tidal field. This is because stellar escape from the outer regions is dominated by tidal forces, while internal dynamics play a secondary role. While the predicted stream mass may be underestimated due to fewer particles, the spatial morphology, kinematics, and radial distributions are robustly reproduced.
Moreover, the simulated cluster's density profile and structure parameters, such as concentration parameter c, $r_h$/$r_t$, are set to be consistent with the real cluster. 
Additionally the simulated cluster is expected to have a smaller velocity dispersion and more rapid spatial mixing than the real cluster, and the escape rate scales inversely with the cluster dissolution timescale. We discuss how these discrepancies between the simulated and real clusters affect the resulting radial distributions in Section~\ref{sect:simu result}.

In this work, we simulate the dynamical evolution of the cluster over the last 800 Myr. This choice is motivated by two considerations. First, we performed a series of test simulations initialized at different look-back times (500, 700, and 1000 Myr ago). We find that stars stripped from the model cluster within the last 800 Myr are sufficient to reproduce the full spatial extent of the observed Fimbulthul stream, which is the focus of this work in terms of its spatial properties and the dynamical imprint of the progenitor. Stars that escaped at earlier times are dispersed over a much more extended region and therefore do not constitute the main body of the Fimbulthul stream.

Second, the relaxation time of the simulated cluster is significantly longer than the simulation duration. Owing to the mass difference between the simulated cluster and the real cluster, their relaxation and dissolution timescales differ substantially. Using the formulae given by \citet{baumgardt2003dynamical}, the two-body relaxation time can be approximated as
\begin{equation}
T_{rh} \sim \frac{N^{1/2} \, r_h^{3/2}}{\langle m \rangle^{1/2} \, G^{1/2} \, \ln(\gamma N)},
\label{eq:trh}
\end{equation}
and the dissolution time is estimated to be:
\begin{equation}
\frac{T_{\mathrm{diss}}}{\mathrm{Myr}} = \beta 
\left( \frac{N}{\ln(0.02N)} \right)^{x}
\left( \frac{R_G}{\mathrm{kpc}} \right)
\left( \frac{V_G}{220\,\mathrm{km\,s^{-1}}} \right)^{-1}
(1 - e),
\label{eq:tdiss}
\end{equation}
where $r_h$ is the half-mass radius, $N$ is the number of stars, $\langle m \rangle$ is the mean stellar mass, $G$ is the gravitational constant, $\gamma$ is the Coulomb logarithm, $R_G$ is the galactocentric distance, $V_G$ is the circular velocity of the Galaxy, and $\beta$ and $x$ depend on the cluster's initial concentration; the cluster's orbit has eccentricity $e$.
Using Eqs.~(\ref{eq:trh}) and (\ref{eq:tdiss}), we estimate that the simulated cluster has a relaxation time of approximately 5~Gyr and a dissolution timescale of about 10~Gyr, whereas the corresponding values for the real cluster are $\gtrsim$12~Gyr \citep[estimated to be $\sim$22~Gyr;][]{2018MNRAS.478.1520B} and $\sim$83~Gyr, respectively.
To facilitate a comparison of their evolutionary stages, we evaluate the dynamical age as the ratio of the cluster’s evolution time to its relaxation time. The simulation duration of 800~Myr therefore corresponds to $\sim0.16\,T_{\rm rh}$ for the simulated cluster (with $T_{\rm rh} \approx 5$~Gyr) and $\sim0.07\,T_{\rm rh}$ for the real cluster (with $T_{\rm rh} \gtrsim 12$~Gyr). 
Since $t \ll 0.2\,T_{\rm rh}$ in both cases, the internal dynamical structure of the cluster (e.g., the degree of spatial mixing and radial gradients) is not expected to evolve significantly during this period. The threshold ($0.2\,T_{\rm rh}$) is chosen based on the work of \citet{Vesperini2013}, who show that radial distributions at $\sim0.2\,T_{\rm rh}$ and $\sim1\,T_{\rm rh}$ are nearly indistinguishable, with substantial differences emerging only after $\sim3\,T_{\rm rh}$.

We also cautiously consider the role of internal dynamical structure, in particular mass segregation. According to \citet{2013MNRAS.435.3272T}, the mass segregation parameter for stars in the mass range $0.5$--$0.8M_{\odot}$ in $\omega$~Centauri is modest, with values of $-0.19$ in the core and approximately $-0.01$ at the half-mass radius. The near-zero value at the half-mass radius suggests that, despite its old age, $\omega$~Centauri exhibits only weak mass segregation in the regions from which a large fraction of escaping stars are expected to originate.
In our analysis, the stellar masses of the cluster sample span $0.7$--$0.9M_{\odot}$, while the stream sample covers $0.7$--$0.85M_{\odot}$. Within this relatively narrow and overlapping mass range, we therefore do not expect strong mass-dependent biases in the stripped population.
Nevertheless, we note that lower-mass stars outside our selection may still preferentially escape and remain observationally undetected, and thus a residual mass dependence in the tidal debris cannot be entirely ruled out. Future work will require long-term simulations of a more massive cluster to investigate the evolution of this early debris and its connection to the cluster’s multiple stellar populations.


\subsection{Initial condition }
We simulated the cluster dynamical evolution --- starting from 0.8 Gyr ago to present day --- to reproduce the observed stream structure. In this simulation, we applied \texttt{MWPotential2014} potential, which is modeled by combining the GALPY code \citep{bovy2015galpy} with PeTar. 

\textbf{Step one} : We performed a backward orbit integration in a Galactocentric reference frame, in which the Sun is at $(x_\odot, y_\odot, z_\odot) = (-8.0, 0., 0.015) \,\mathrm{kpc}$, with velocity $(U_\odot, V_\odot, W_\odot) = (10, 235, 7) \,\mathrm{km\,s^{-1}}$. The determined initial position of the cluster center is $(x,y,z)=(4.0, 0.3, 1.6) \,\mathrm{kpc}$ and the initial velocity is  $(V_x, V_y, V_z) = (-186.9, 129.4, 16.3) \,\mathrm{km\,s^{-1}}$ 0.8 Gyr ago.

\textbf{Step two} : To determine the stellar mass function of the cluster at a lookback time of $\sim 800$~Myr, we first generated a model cluster using \texttt{mcluster}. The initial model consisted of $N = 3 \times 10^5$ particles with a total mass of $M = 1.78 \times 10^5\,M_{\odot}$, following a King density profile with a dimensionless central potential of $W_0=8$. We adopted the \citet{kroupa2001variation} initial mass function (IMF) spanning the mass range $0.08 \le m/M_{\odot} \le 150$. To obtain the evolved masses of individual stars, the sample was evolved to an age of $11.2$~Gyr using the \texttt{petar.bse} module. This module utilizes a significantly updated implementation of the \texttt{BSE} package \citep{2000MNRAS.315..543H, 2002MNRAS.329..897H}, incorporating state-of-the-art stellar evolutionary recipes. The evolution was performed in isolation with the metallicity set to $Z = 0.000156$, corresponding to the median metallicity of $\omega$~Centauri. We note, however, that $\omega$~Centauri exhibits a broad metallicity distribution, and the adoption of a single representative metallicity is a simplifying assumption in our modeling.

\textbf{Step three}: We removed unbound high-velocity stars and rescaled the velocities to enforce virial equilibrium. The initial conditions assume negligible mass segregation, motivated by observations of the internal dynamics of $\omega$~Centauri \citep{2013MNRAS.435.3272T}.



\begin{figure*}
\centering
\includegraphics[width=\textwidth]{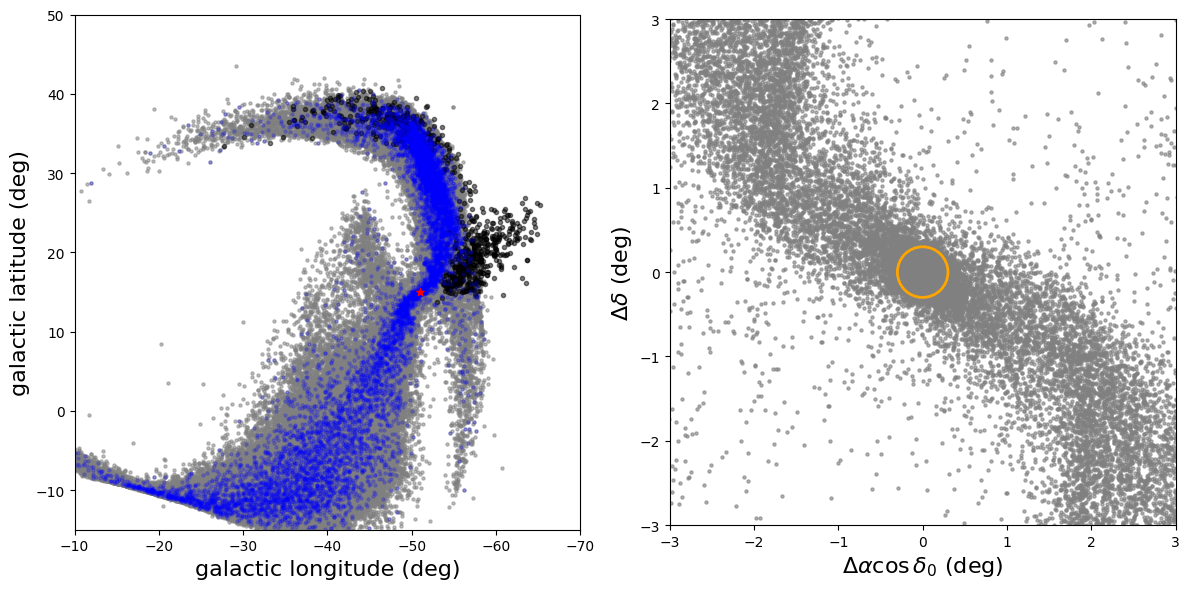}
\caption{Position of the simulated particles. In the left panel, all particles are shown as grey dots, the selected particles as blue dots, the observed stars as dark dots, and the cluster center as a red star. The right panel provides a zoom–in on the tidal stream, illustrating the cluster structure and its tidal extension; the yellow circle indicates the main body of the simulated cluster, with a radius of 20 arcmin.}
\label{fig:pos.particle}
\end{figure*}

\begin{figure} 
\centering
\includegraphics[width=\columnwidth]{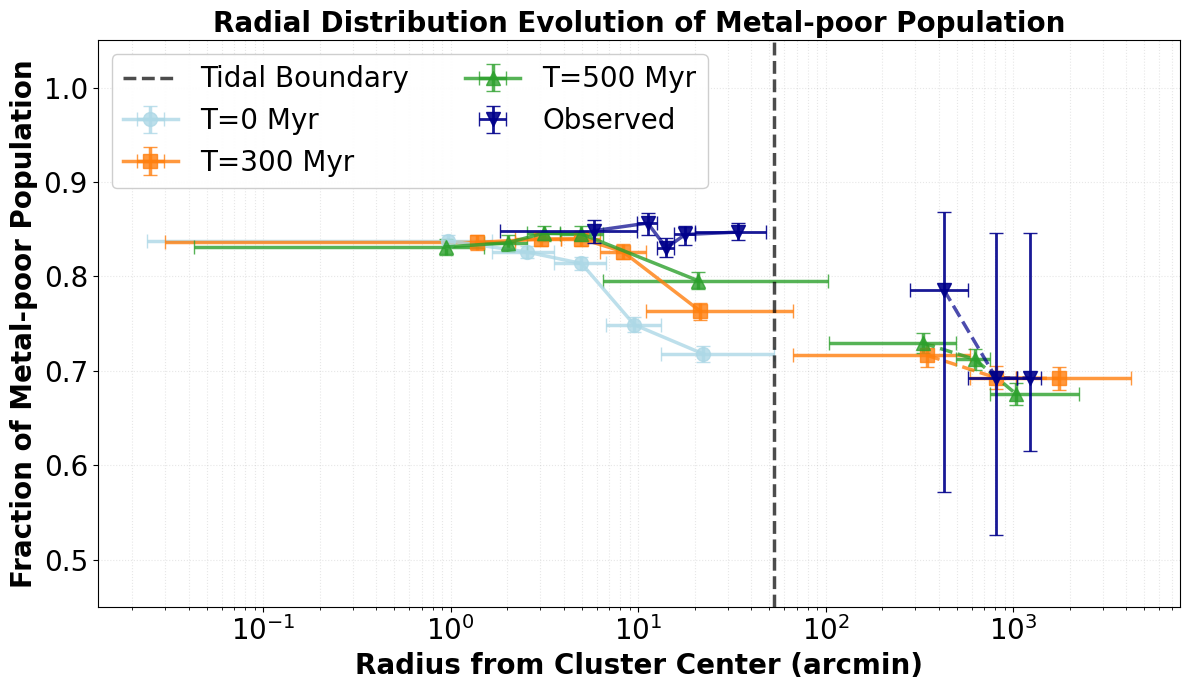}  
\caption{Radial distributions at four evolutionary stages: $t=0$~Myr (light blue), $300$~Myr (orange), $500$~Myr (green), and the present day at $800$~Myr (dark blue). No tidal stream is present at $t=0$, whereas the outer regions at later times consist of escaping stars. The initial $r_{\rm t}$ (at $t=0$) is indicated as a fixed boundary (black dashed line) to distinguish the stream, noting that the intrinsic $r_{\rm t}$ evolves with time.
}
\label{fig:sim_radial}
\end{figure} 

\begin{figure} 
\centering
\includegraphics[width=\columnwidth]{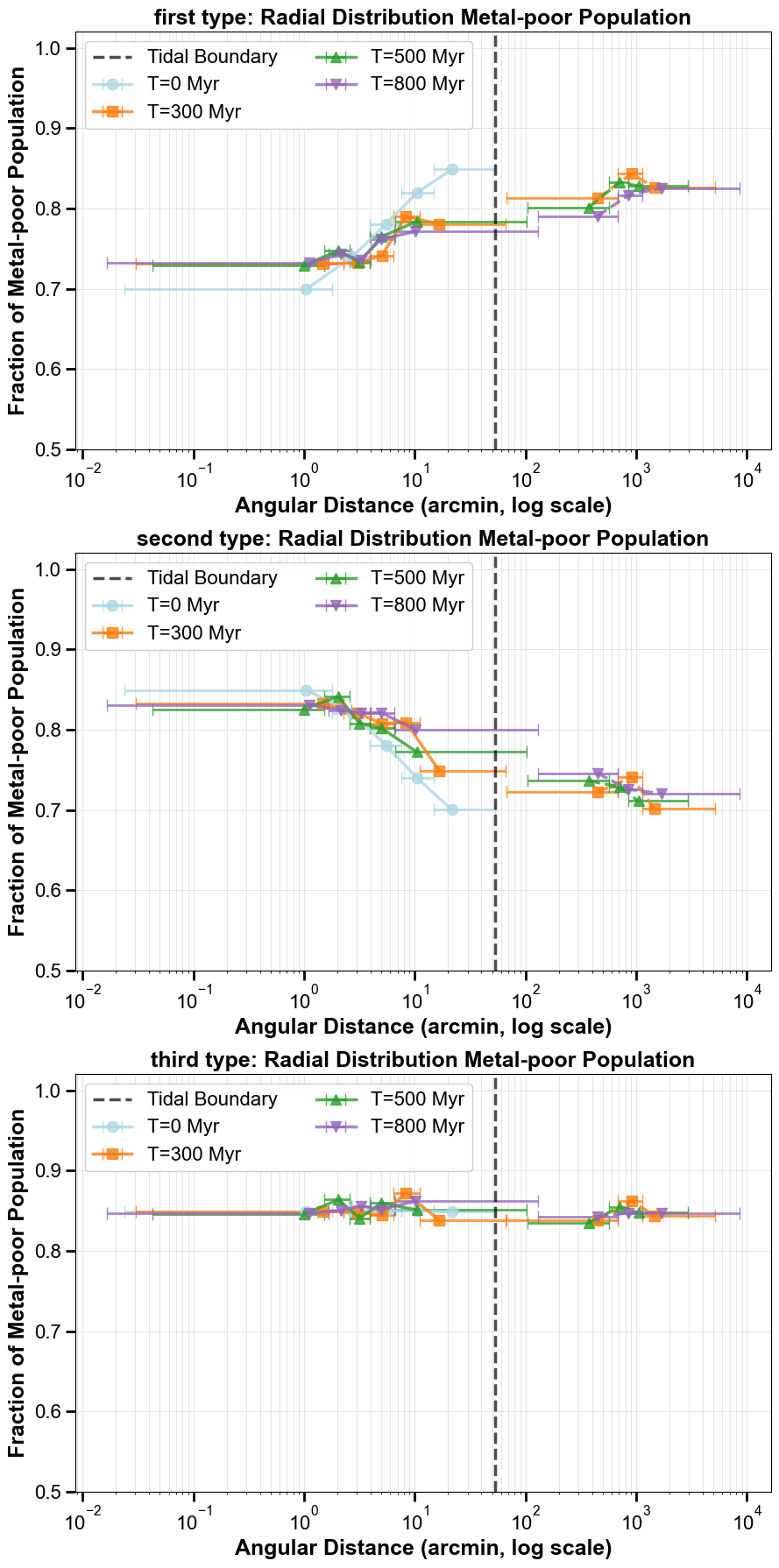}  
\caption{The three panels show the radial distributions at the same evolutionary stages but with different initial configurations: increasing, flat, and decreasing patterns. The initial radial distribution exhibited by fractions of metal-poor population in five bins, markered as light blue dots.The black dashed line denotes the tidal boundary, as in Fig.\ref{fig:sim_radial}}
\label{fig:sim_radial_3type}
\end{figure} 

\subsection{Simulation result}
\label{sect:simu result}
The present-day spatial distribution of the simulated particles in Galactic coordinates is shown in the left panel of Figure~\ref{fig:pos.particle}. The right panel shows the corresponding distribution in equatorial coordinates, exhibited relative to the cluster center. Although the simulated stream does not fully reproduce the observed stream morphology, the possible reasons for this mismatch will be discussed in Section~\ref{sect:discuss_caveats}. 

We compared the backward evolution of radial distribution with the forward evolution of that one. One was constructed to reproduce the present radial distribution for simulated cluster inferred from observations and was then traced backward in time, while the other started from a specific initial radial distribution to obtain its radial evolution at different evolutionary phases. Comparing the backward and forward evolution of the radial distribution is to assess the reliability of the initial radial distribution inferred from the backward integration.

To describe the evolution of radial distribution of different populations with different metallicity, we analysed how the fraction of metal-poor population changed with time. According to the observed sample, we select a subset of stream particles with stellar masses between 0.7 and 0.85~$M_\odot$, located at angular distances of 270–1800 arcmin, as well as cluster particles within 50 arcmin with masses between 0.7 and 0.9~$M_\odot$, as determined from isochrone fitting of observed stars. This selection ensures that the simulated sample roughly matches the observed stellar populations in both spatial extent and mass range.
Then, we grouped the cluster stars into five bins and the stream stars into three bins according to their angular distance from the cluster center. The simulation particles in each bin were further divided into metal-poor and metal-rich stars based on the observed fraction of metal-poor stars. In this way, the simulated particles reproduce the same radial distribution as observed, which was exhibited as blackblue markers in Figure~\ref{fig:sim_radial}. 

We then back-trace the evolution of the radial distributions at different epochs, which provides valuable insights into the dynamical evolution of different stellar populations and their initial spatial distributions within the cluster. Figure~\ref{fig:sim_radial} shows the radial distributions at several evolutionary stages. Within a radius of 5~arcmin, the stellar fractions exhibit only minor changes with time, whereas in the outer regions of the cluster, extending into the tidal stream, the fraction increases by approximately 0.1~dex as the cluster evolves. The light-blue points represent the initial radial distribution of the cluster, which is nearly flat from the center out to $\sim$5~arcmin, with only a mild variation of $\sim$0.1~dex at larger radii.
We note that when fainter stream stars are included, the inferred metal-poor fraction in the stream becomes comparable to that of the cluster, and the resulting radial gradient remains nearly flat throughout the simulated evolution as shown in Figure \ref{fig:A3}.

To assess the reliability of the initial radial distribution inferred from the backward integration, we initialized three radial distributions corresponding to increasing, decreasing and flat trends in the fraction of the metal-poor population (Figure~\ref{fig:sim_radial_3type}). In the first distribution, the fraction increases linearly from 0.7 to 0.85 with radius across five equal-number bins; in the second distribution, it decreases linearly from 0.85 to 0.7; and it remains constant at 0.85 in the third distribution. For each initial radial distribution, we performed twenty random relabelings of the simulation particles to assess the impact of stochasticity on the resulting profiles. We find that the uncertainty in the derived fractions due to random labeling is at the level of $\sim 0.5\%$, and that the resulting radial distributions are insensitive to stochastic effects.

Our simulations show that two-body relaxation and the external tidal field act to flatten the initial radial gradients in the fraction of the metal-poor population within the bound cluster. After 0.16 relaxation times (800/5000~Myr for the simulated cluster), the differences in the metal-poor fraction among cluster stars decrease from $\sim 0.15$~dex to approximately $\sim 0.05$~dex for both the initially increasing and decreasing distributions.
However, when considering the entire system including the tidal tails, a radial gradient of $\sim 0.15$~dex from the cluster center to the tidal tail is still preserved over time. This indicates that the regions beyond the $r_{\rm t}$ (i.e., the tidal tails) contribute significantly to the overall radial distribution and act as reservoirs of the initial structural information.

Even over longer evolutionary timescales of $\sim$0.16–1 relaxation times, the initial radial gradient is not expected to change significantly. This is supported by previous studies \citep{Vesperini2013}, which show that the radial distributions after $\sim$0.2 and $\sim$1 relaxation times are nearly indistinguishable, whereas substantial differences among stellar populations only emerge after $\sim$3 relaxation times.
Therefore, we infer that the radial distribution of the cluster at 800~Myr ago is likely a good approximation to its primordial radial distribution, and that the present-day radial structure still retains memory of the initial distribution established at the formation epoch of the cluster ($\sim$12~Gyr ago), corresponding to less than one relaxation time for $\omega$~Centauri.

Current data do not allow us to distinguish between an initially radially flat metal-poor fraction and one with a mild decrease ($\sim$0.15~dex). However, from the evolution of the radial profile we infer that the initial gradient is linked to the difference in the metal-poor fraction between the cluster and the stream. Further observational and theoretical studies of multiple populations in streams are therefore needed to advance our understanding of multiple-population formation and evolution in clusters.


\section{Discussion}  
\label{sect:dis}

\subsection{Caveats}
\label{sect:discuss_caveats}
Compared with the simulated distribution, the observed stream structure doesn't directly connect with the progenitor's main body. It might be attributed to the low Galactic latitude of $\omega$ Centauri ($|b|<15^{\circ}$), where high source density and extinction render it very challenging to detect streams. An alternative and physically well-motivated explanation is mass segregation. In dynamically evolved clusters, low-mass and low-luminosity stars are preferentially stripped at late times and are expected to dominate the tidal debris. Since our cluster (0.7–0.9 $M_\odot$) and stream (0.7–0.85 $M_\odot$) samples do not include the lowest-mass stars, the observed stream likely traces only the higher-mass component of the stripped population. Consequently, a significant fraction of the tidal debris may remain undetected, naturally weakening the observed spatial continuity between the stream and its progenitor.

In addition, the observed structure of the Fimbulthul stream exhibits stronger bending than that seen in the simulated particle distributions. Similar discrepancies are also present in the simulations of \citet{ibata2019identification}. In both cases, the simulations adopt scaled representations of cluster evolution rather than fully self-consistent, one-to-one realizations, which may partly account for the observed differences.
Moreover, differences in the adopted Galactic potential models themselves can lead to significant variations in stream morphology, including enhanced bending, fanning, and the formation of gaps. Incorporating the effects of dark matter substructures \citep{yavetz2023stream} or adopting alternative Galactic potential models \citep{2017NatAs...1..633P} may therefore produce stream morphologies that more closely reproduce these observed features. Another possibility is that close passages of dense stellar systems, such as globular clusters, could induce additional perturbations to the stream \citep{2025A&A.Ferrone}.

Nevertheless, our simulation results should be interpreted with caution. This dynamical model cannot fully capture the complex evolutionary history of $\omega$ Centauri, which involves multiple stellar generations and interactions between the progenitor dwarf galaxy and the Milky Way \citep{mason2025chemical, jofre2025studying, gonzalez2025chronology}.

\subsection{Formation scenario }
By combining our results with existing literature, a coherent picture emerges for the spatial and chemical structure of $\omega$ Centauri's stellar populations. The key observational constraints are: 1) the ratio of metal-poor to metal-rich stars remains roughly constant \citep[this work;][]{Nitschai2024oMEGACatIM, clontz2025}, 2) The proton-normal (1P) population is significantly older than the proton-enriched (2P) population \citep{clontz2024omegacat,clontz2025}, and 3) the 2P population is more centrally concentrated than the 1P population \citep{2025arXiv250916719D,2009A&A...507.1393B, 2024A&A...688A.180S}. Given the cluster's dynamical evolution, the initial central concentration of the proton-enriched population was likely even more pronounced.

We find the formation scenarios proposed by \citet{2025arXiv250916719D} and \citet{clontz2025} to be generally compelling. In this context, we elaborate on the spatial distribution of the distinct stellar populations: 1P metal-poor (1P-MP); 2P metal-poor (2P-MP); 1P metal-rich (1P-MR); and 2P metal-rich (2P-MR). 
The sequence likely began with the formation of the 1P-MP population from a primordial gas cloud. 
The Subsequent onset of core-collapse supernovae (CCSNe) injected high-velocity shocks ($> 10^4$ km/s) after a few tens of Myr. Due to $\omega$ Centauri's substantial potential well—consistent with a nuclear star cluster—this feedback was insufficient to completely expel the remaining gas. Instead, the CCSNe winds enriched the interstellar medium and triggered a new wave of star formation. Because CCSNe winds are faster and more energetic than typical stellar outflows, the resulting 1P-MR population formed in a less centralized manner, explaining its observed flat radial gradient.
Concurrently, proton-rich material was injected by other astrophysical pollutors, e.g., very massive stars \citep{Vink2018}, extremely massive stars \citep{Gieles2025}, or massive/intermediate-mass AGB stars \citep{Ventura2013}. Their slower winds ($\sim 10^2-10^3$ km/s) allowed the material to sink deep into the cluster's potential well. This process led to the formation of the more centrally concentrated 2P population. The disruptive shocks from the earlier CCSNe likely delayed the accumulation of this proton-rich gas, requiring up to $\sim 1$ Gyr to reach critical density \citep{clontz2025}. 
The mixing of this proton-rich gas with pristine gas resulted in the 2P-MP population. In contrast, its mixing with the CCSNe-enriched gas produced the 2P-MR population. In both cases, the formation mechanism naturally explains why the 2P populations are more centrally concentrated than their 1P counterparts. This dynamical picture provides a robust explanation for the roughly constant metal-poor fraction observed globally. However, depending on the precise dominance of the central 2P-MP population, a secondary effect may emerge: a mild radial decline in the metal-poor fraction. This possibility highlights the scenario's capacity to explain both the general flatness and subtle second-order trends



\vspace{0.8cm} 
\section{Summary}
\label{sect:sum}

The spatial distribution of multiple stellar populations provides key insights into the formation and evolutionary history of globular clusters.
In this work, we investigated the radial distributions of stellar populations with different metallicities within $\omega$ Centauri and its tidal stream. This work is part of our ongoing research project, ``Scrutinizing {\bf GA}laxy-{\bf ST}a{\bf R} cluster coevoluti{\bf ON} with chem{\bf O}dyna{\bf MI}cs ({\bf GASTRONOMI})'', which leverages multi-wavelength photometric and spectroscopic data to unravel the coevolutionary relationships between the MW, its satellite dwarf galaxies, and their star clusters. 

We find no significant radial gradient in the population ratios within the cluster. Although metal-rich stars appear to be slightly more extended along the tidal stream, this trend is not statistically significant once the observational uncertainties are taken into account. Overall, the population ratios are consistent with being nearly constant from the cluster center to the tidal tails.
By simulating the dynamical evolution of the cluster, we infer that the initial radial distributions of different stellar populations were likely close to flat. 
For a cluster evolving for less than one relaxation time, the radial gradients in population fractions are not expected to change dramatically compared to their initial distributions. Nevertheless, the combined effects of two-body relaxation and tidal stripping already induce spatial mixing among different stellar populations, leading to a tendency for the population fractions to become progressively flatter with radius within cluster.
Our results characterize how the radial profiles of multiple metallic populations are shaped by the combined effects of internal relaxation and the external tidal field, providing new insights into the formation and dynamical evolution of multiple metallic populations in this unique system.

As relics of globular clusters and dwarf galaxies, the multiple populations traced by tidal streams provide key clues to the assembly histories of globular clusters and their parent systems \citep{2022MNRAS.515.5802B, usman2024multiple, casey2021signature, simpson2020galah}. In the future, Fernandez-Trincado et al. (in prep.) will provide a new high-resolution spectrum analysis for Fimbulthul stream soon. High-resolution spectroscopy of faint stream stars, combined with dynamical simulations, is essential to trace their chemical patterns and link them to progenitor globular clusters or dwarf galaxies.

\begin{acknowledgements}
S.Z. and B.T. gratefully acknowledge support from the National Natural Science Foundation of China through grants NOs. 12233013 and 12473035, China Manned Space Project under grant NO. CMS-CSST-2025-A13 and CMS-CSST-2021-A08, the Fundamental Research Funds for the Central Universities, Sun Yat-sen University (24qnpy121).
L.W. thanks the support from the National Natural Science Foundation of China through grant 12573041 and 21BAA00619, the High-level Youth Talent Project (Provincial Financial Allocation) through the grant 2023HYSPT0706, the one-hundred-talent project of Sun Yat-sen University, the Fundamental Research Funds for the Central Universities, Sun Yat-sen University (2025QNPY04). The authors acknowledge the Beijing Beilong Super Cloud Computing Co., Ltd for providing HPC resources that have contributed to the research results reported within this paper.URL: http://www.blsc.cn/.
J.G.F-T gratefully acknowledges the support provided by ANID Fondecyt Regular No. 1260371, ANID Fondecyt Postdoc No. 3230001 (Sponsoring researcher), the Joint Committee ESO-Government of Chile under the agreement 2023 ORP 062/2023 and the support of the Doctoral Program in Artificial Intelligence, DISC-UCN.
\end{acknowledgements}


\bibliography{sample701}{}
\bibliographystyle{aasjournalv7}

\section*{APPENDIX}
\addcontentsline{toc}{section}{APPENDIX}

\setcounter{figure}{0}
\renewcommand{\thefigure}{A\arabic{figure}}
\setcounter{table}{0}
\renewcommand{\thetable}{A\arabic{table}}
\vspace{0.3cm}
\begin{figure*}[h!]
    \centering
    \includegraphics[width=0.55\textwidth]{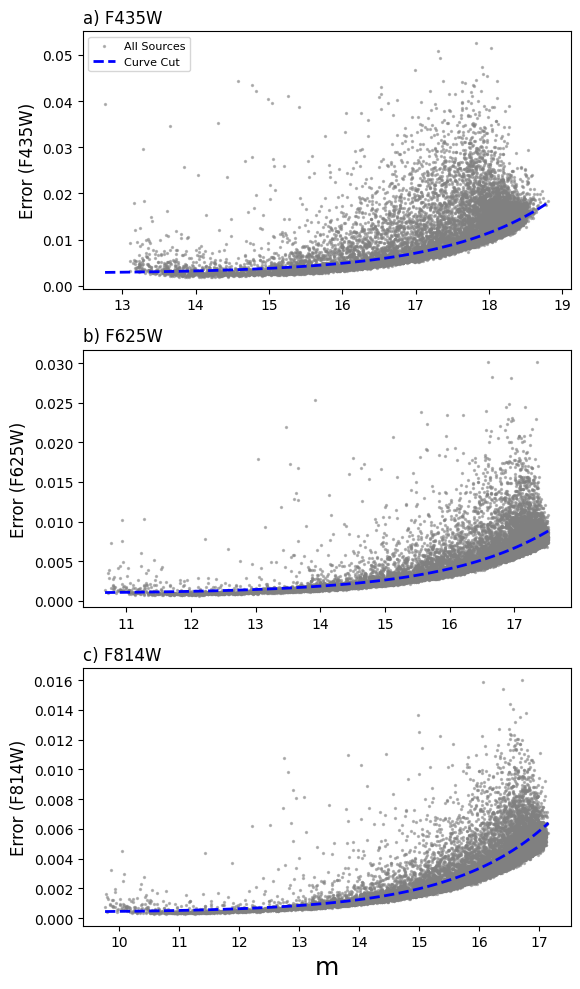} 
    \caption{
        Photometric error ($\sigma_{\text{m}}$) as a function of magnitude for the F435W, F625W, and F814W filters.
        The grey points represent the initial sample before quality cuts.
        The blue dashed curves indicate the upper boundaries used to define our high-accuracy sample.
        Following the strategy of \citet{2005MNRAS.357..265S}, the behavior of photometric errors has been modelled as a function of magnitude using analytical exponential functions, adopting the curves that provide $\sigma_{\text{F435W}} = 0.01$ at $m=18$, and $\sigma = 0.006$ at $m=17$ for both F625W and F814W.
        Stars lying above these boundaries were rejected from the final catalog.
        (See also \citealt{2007Sesar} for the theoretical noise model).
    }
    \label{fig:error_cuts}
\end{figure*}

\vspace{1.0cm}

\begin{figure*}[h!]
    \centering
    \includegraphics[width=0.95\textwidth]{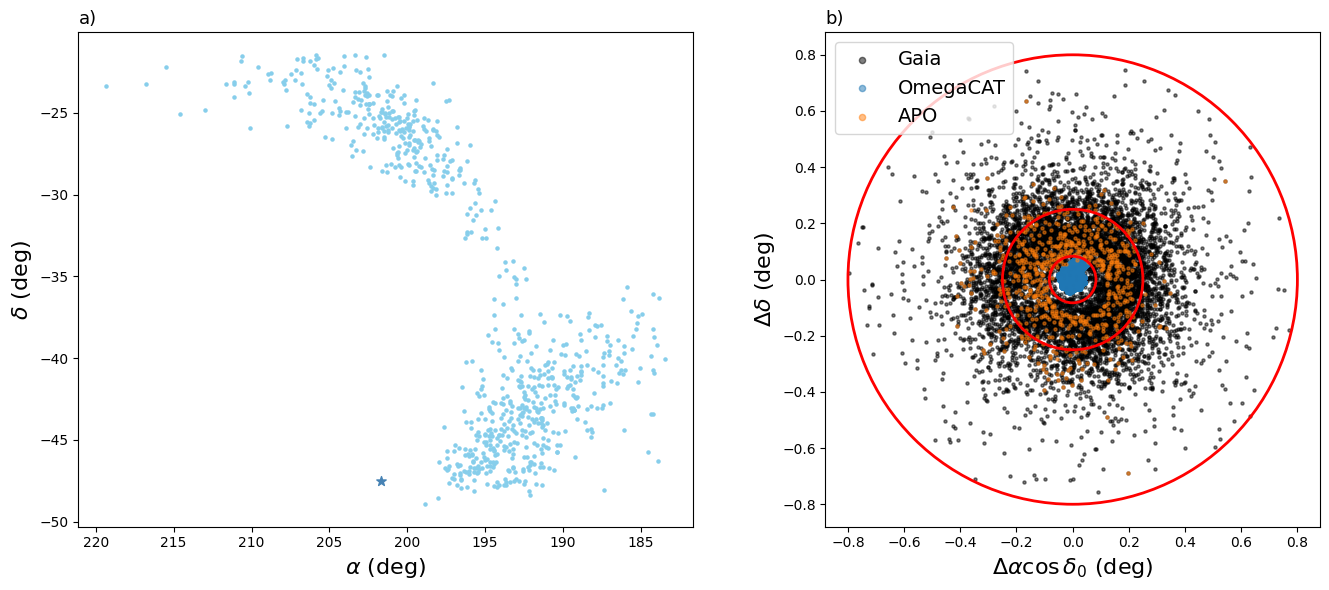}
    \caption{
        Spatial distribution of cluster members and stream stars.
        \textit{Left panel:} Stream members cross-matched to Gaia XP, with the steel-blue asterisk indicating the position of the cluster center.
        \textit{Right panel:} Cluster members selected from three sources: blue dots in the core represent members with metallicities from MUSE \citep{Nitschai2024oMEGACatIM}; orange dots denote members with metallicities from APOGEE \citep{2021MNRAS.505.1645M}; black dots in the outer region (approximately 5$-$50 arcmin) represent members selected from Gaia DR3.
        The red circles indicate the radii corresponding to one half-light radius, three half-light radii, and $r_{\rm t}$.
    }
    \label{fig:pos}
\end{figure*}

\begin{figure*}[h!]
    \centering
    \includegraphics[width=0.95\textwidth]{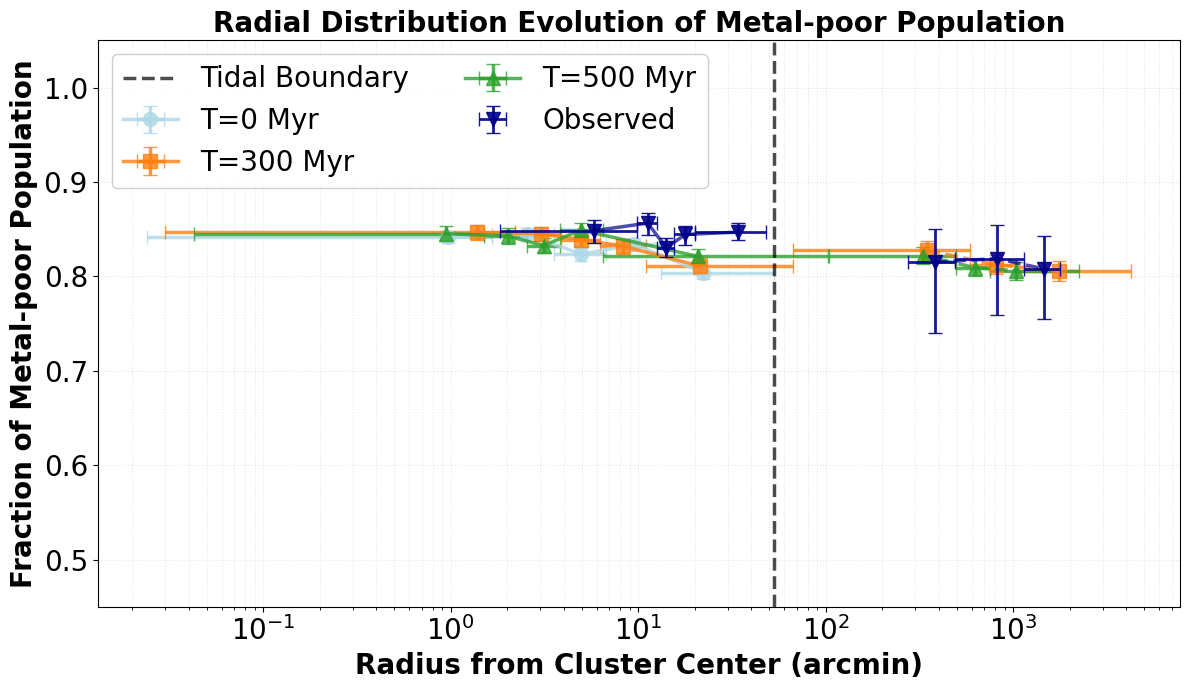}
    \caption{
        Radial distribution evolution based on sample including fainter stream stars.  
    }
    \label{fig:A3}
\end{figure*}

\FloatBarrier



\end{document}